\documentstyle[preprint,pra,aps]{revtex}

\def\ket#1{|#1\rangle} \def\bra#1{\langle#1|}
\def\av#1{\langle#1\rangle}

\tightenlines

\begin{document} 
\draft 
\title{Two-state system driven by imperfect $\pi$ pulses: an estimate of the error accumulation in
bang-bang control methods} 
\author{Julio Gea-Banacloche} 
\address{Department of
Physics, University of Arkansas, Fayetteville, AR 72701} 
\date{\today}
\maketitle 
\begin{abstract}  The evolution of a two-state system driven by a sequence of imperfect $\pi$ pulses
(with random phase or amplitude errors) is calculated.  The resulting decreased fidelity is used to
derive a plausible limit on the performance of "bang-bang" control methods for the suppression of
decoherence.  
\end{abstract}

\narrowtext

\section{Introduction and model}
A number of dynamical methods for the suppression of decoherence in quantum systems have recently been
proposed \cite{viola1,ban,viola2,vitali,agarwal,duan}.  While some of these are quite sophisticated, the
simplest such approach, sometimes called
``bang-bang'' control \cite{viola1,ban}, is a rather straightforward idea which could, in principle, be
applied to any
two-state system interacting with an environment which has a {\it finite} (nonzero) correlation time
$\tau_c$.  (See \cite{kwiat} for a recent proof-of-principle experimental demonstration.)  The idea is to
``flip'' rapidly (faster than $\tau_c$) the state of the system, back and forth, in such a way that the
environment's unwanted influence on the system is constantly being undone by the environment itself.

For instance, for a system to environment coupling which involves only the operator $\sigma_z$ (pure
phase decoherence), due to a term such as
\begin{equation}
\sum_k \hbar \sigma_z \left(g_k b_k^\dagger + g_k^\ast b_k \right)
\label{zero}
\end{equation}
in the  Hamiltonian (where the boson operators $b_k$ represent modes of the environment), a sequence of
very short $\pi$ pulses, which rotate the system's pseudospin by $180^\circ$ around the $x$ axis and
hence change the sign of $\sigma_z$, would result in an evolution in which the sign of (\ref{zero})
changes form one instant to the next, and its effect therefore averages to zero.  

The purpose of this note is to consider the constraints which imperfections in the $\pi$ pulses
place on the successful implementation of such a dynamical decoupling strategy.  The approach is simply
to consider the evolution of a two-state system subject to a series of imperfect $pi$ pulses, and to
calculate the total error (as measured by the state's fidelity) introduced, on the average, after $N$
such pulses.  The average considered here is over all possible initial states of the system, as well as
over the distribution of the random errors in the $\pi$ pulses.

I work in the interaction picture and assume that the central frequency of the pulses is exactly tuned to
the resonance frequency of the transition between the two states of the system.  The rotating-wave
approximation is also assumed to apply.  The Hamiltonian describing the interaction of the system with a
single pulse is then
\begin{equation}
H = \hbar g \left({\cal E}(t)\sigma_+ + {\cal E}^\ast(t) \sigma_-\right)
\label{one}
\end{equation}
where $\sigma_+ = \ket{+}\bra{-}$,  $\sigma_- = \ket{-}\bra{+}$, and $\ket+$ and $\ket-$ are the
eigenstates of $\sigma_z$.  The coupling constant $g$ can be taken to be real without loss of generality. 
In general, the Hamiltonian (\ref{one}) does not commute with itself at different times.  To simplify
matters, I shall take the classical field amplitude to be of the form ${\cal E}(t) = E(t) e^{i\varphi}$,
where $E(t)$ is real and $\varphi$ is a constant, determined essentially by the timing of the pulse. 
Then, the evolution operator corresponding to (\ref{one}) is, in the  \{$\ket+$,$\ket-$\} basis, 
\begin{eqnarray}
U(t) &&= \exp\left[-{i\over\hbar} \int_0^t H(t^\prime) dt^\prime \right] \nonumber \\
&&= \exp\left[-i\left(\cos\varphi \sigma_x - \sin\varphi\sigma_y\right)\int_0^t gE(t^\prime)
dt^\prime\right]
\label{two}
\end{eqnarray}
If the duration of the pulse is $t_p$, define the angle $\Theta \equiv 2\int_0^{t_p} gE(t^\prime)
dt^\prime$.  In pseudospin terms, the operator $U(t_p)$ represents a rotation of the spin by an
angle $\Theta$ about an axis in the $x-y$ plane which makes an angle $\varphi$ with the $x$ axis.  If
$\Theta=\pi$, one has a $\pi$ pulse.

In this paper I consider the state resulting from a sequence of ``imperfect'' $\pi$ pulses. 
Imperfections may arise from either the phase $\varphi$, which controls the axis of the rotation, and
which would be sensitive to errors in the timing of the pulses, or from errors in the amplitude and/or
duration of the pulse, which would make $\theta \ne \pi$.  Amplitude errors are considered in Section 2,
phase errors in Section 3, and the results, and their possible consequences for ``bang-bang'' control,
are discussed in Section 4.

\section{Amplitude errors} 

In this Section I shall assume that $\varphi = 0$, in Eq.~(\ref{two}), for all the pulses in the
sequence.  A full spin cycle consists of two $\pi$ pulses (separated by a short time of ``free''
evolution), so I shall consider a total of $2N$ pulses, where $N$ is the total number of cycles.  I will
take the ``pulse area'' $\Theta$ to be of the form $\Theta = \pi + \epsilon_i$, where $\epsilon_i$ is a
small number which varies randomly from one pulse to the next.  All the $\epsilon_i$ are assumed to be
distributed in a Gaussian way, that is, according to the following probability distribution:
\begin{equation}
P(\epsilon_i) = {1\over\sqrt{2 \pi}\Delta} e^{-\epsilon_i^2/2\Delta^2}
\label{three}
\end{equation}
with standard deviation $\Delta$.

With this notation the evolution operator (\ref{two}) for the $i$-th pulse is simply
\begin{equation}
U_i = e^{-i(\pi/2+\epsilon_i/2)\sigma_x}
\label{four}
\end{equation}
and the result of $2N$ consecutive pulses is given by
\begin{eqnarray}
U_{total} &&=  e^{-i(N\pi+\epsilon_1/2+\ldots\epsilon_{2N}/2)\sigma_x} \nonumber \\
&&\equiv e^{-i(N\pi+\epsilon/2)\sigma_x} \nonumber \\
&&= (-1)^N \left[\cos{\epsilon\over 2} -i\sin{\epsilon\over 2} \sigma_x \right]
\label{five}
\end{eqnarray}
where the random variable $\epsilon = \epsilon_1+\ldots\epsilon_{2N}$ is also distributed in a Gaussian
way, and has standard deviation $\sqrt{2N}\Delta$.
 
A general initial state of the two-level system can be written as
\begin{equation}
\ket{\psi_0} = \cos{\theta\over 2}\ket+ + e^{i\phi}\sin{\theta\over 2}\ket-
\label{six}
\end{equation}
In this state the spin is aligned along an axis of colatitude $\theta$ and azimuth $\phi$.  The effect of
(\ref{five}) on the state (\ref{six}) is 
\begin{equation}
U_{total}\ket{\psi_0} = (-1)^N\left[\left(\cos{\epsilon\over 2}\cos{\theta\over 2}-ie^{i\phi}
\sin{\epsilon\over 2}\sin{\theta\over 2}\right)\ket+ + \left(\cos{\epsilon\over 2}\sin{\theta\over
2}e^{i\phi}-i \sin{\epsilon\over 2}\cos{\theta\over 2}\right)\ket- \right]
\label{seven}
\end{equation}
The fidelity, $\cal F$, of the state after all the pulses is given by ${\cal F} = \left|\bra{\psi_0}
U_{total}\ket{\psi_0}\right|^2$, and it equals
\begin{equation}
{\cal F} = \cos^2{\epsilon\over 2}+\sin^2{\epsilon\over 2} \sin^2\theta\cos^2\phi
\label{eight}
\end{equation}
To calculate an average fidelity, Eq.~(\ref{eight}) may be integrated over $\theta$ and $\phi$, assuming
a uniform distribution of initial states over the whole Bloch sphere (that is, with a density $\sin\theta
d\theta d\phi/4\pi$), and over $\epsilon$ with the appropriate probability distribution.  Before
proceeding to do so, however, note that $\sin\theta\cos\phi$ is just $x$ in Cartesian coordinates; hence
a reparametrization of the Bloch sphere that exchanges $x$ and $z$ will turn $\sin\theta\cos\phi$ into
$\cos\theta^\prime$, which makes the integral over $\phi^\prime$ trivial.  The probability distribution
for the final state fidelity can thus be written formally as
\begin{eqnarray}
P({\cal F}) &&= {1\over 2 \Delta} {1\over \sqrt{4 N \pi}} \int_{-\infty}^{\infty}
e^{-\epsilon^2/4N\Delta^2} d\epsilon\int_0^\pi sin\theta^\prime d \theta^\prime \delta\left({\cal F} -
\cos^2{\epsilon\over 2} - \cos^2 \theta^\prime \sin^2{\epsilon\over 2} \right) \nonumber \\
&&= {1\over \Delta} {1\over \sqrt{4 N \pi}} \sum_{n=-\infty}^\infty \int_{-\sqrt{\cal F}}^{\sqrt{\cal F}}
{\exp\left[-{1\over
N\Delta^2}\left(\sin^{-1}\sqrt{1-{\cal F}\over 1 - x^2}-n\pi\right)^2\right]}\,{dx\over \sqrt{(1-{\cal
F})({\cal F}- x^2)}}
\label{nine}
\end{eqnarray}
The last integral may be evaluated numerically for a given value of $N\Delta^2$ and ${\cal F}$ (with
$0<{\cal F}<1$); for reasonable values of $N\Delta^2$, only a few terms in the sum over $n$, around
$n=0$, are necessary.  The results are shown in Figure 1 for $N\Delta^2 = 0.1$, $1$, and $10$,
respectively.  The divergence as ${\cal F} \to 1$ is integrable; the value of $P({\cal F})$ as ${\cal F}
\to 0$ is finite.  The average of ${\cal F}$ can be calculated easily, and the result is
\begin{equation}
\av{{\cal F}} = {2\over 3} + {1\over 3} e^{-N\Delta^2}
\label{ten}
\end{equation}
This shows the expected decrease in fidelity with the number of cycles, but it also shows that for
sufficiently large $N\Delta$ the average fidelity saturates at the value $2/3$.  This may look like a
relatively large number, but one must keep in mind that for a two-level system a totally random mixed
state with density operator $\rho = 1/2$ would still yield a fidelity of $0.5$ when compared to any
initial state.  Thus, in fact, a fidelity of $0.67$ for this system corresponds to an almost random final
state.  

This is illustrated in Figure 2, which shows three simulations of trajectories starting from three
randomly picked states and going through a total of $N_{max}=400$ cycles, with $\Delta$ chosen in each
case so
that $N_{max} \Delta^2 = 0.1$, $1$, and $10$, as in Fig.~1 (the corresponding average fidelities, as
given
by Eq.~(\ref{ten}), are $0.968$, $0.789$ and $0.667$).  Clearly, the trajectory for $\sqrt N_{max} \Delta
= 10$
is all over the place.  (The trajectories shown have been chosen from a relatively large sample of
randomly generated trajectories so as to be as ``average'' as possible, that is, neither unusually
``good'' nor unusually ``bad'' for the particular value of $N_{max} \Delta^2$ that they illustrate.)

In this connection it may be worthwhile to consider the ``worst-case fidelity'' predicted by
Eq.~(\ref{eight}):  it corresponds to initial states having either $\theta = 0, \pi$, or $\phi=\pi/2$,
and, when averaged over $\epsilon$, yields
\begin{equation}
\av{{\cal F}_{min}} = {1\over 2} + {1\over 2} e^{-N\Delta^2}
\label{tenandahalf}
\end{equation}
For large $N\Delta^2$, this gives $1/2$, i.e., the fidelity of the totally random mixed state, as
discussed above.

\section{Phase errors}

In this Section I take the pulse area $\Theta \equiv 2\int_0^{t_p} gE(t^\prime)
dt^\prime$ to be exactly $\pi$, and the phase $\varphi$ in Eq.~(\ref{two}) to be a small
random number, varying from pulse to pulse, with zero average and a Gaussian distribution with standard
deviation $\Delta$.  Then the evolution operator for the $i$-th pulse is
\begin{eqnarray}
U_i &&= e^{-i(\pi/2)(\cos\varphi_i\sigma_x-\sin\varphi_i\sigma_y)} \nonumber \\
&&=-i(\cos\varphi_i\sigma_x-\sin\varphi_i\sigma_y)
\label{eleven}
\end{eqnarray}
and its effect on the general state (\ref{six}) is 
\begin{equation}
U_i \ket{\psi_0} = e^{-i\varphi_i}\left(\cos{\theta\over 2}\ket+ + e^{i(\phi+2\varphi_i)}\sin{\theta\over
2}\ket- \right)
\label{twelve}
\end{equation}
that is, except for an overall phase factor, all that happens is that the angle $\phi$ becomes
$\phi+\varphi_i$.  Therefore, after $N$ cycles totalling $2N$ pulses, the final state will be, up to a
phase factor
\begin{equation}
U_{total} \ket{\psi_0} = \cos{\theta\over 2}\ket+ + e^{i(\phi+\epsilon)}\sin{\theta\over 2}\ket- 
\label{thirteen}
\end{equation}
where $\epsilon = 2(\varphi_1+\ldots+\varphi_{2N})$ is now a random variable with zero average and a
Gaussian distribution with standard deviation $2\sqrt{2 N} \Delta$.  A little trigonometric manipulation
shows that the fidelity of the state (\ref{thirteen}) can be written as
\begin{equation}
{\cal F} = \cos^2{\epsilon\over 2}+\sin^2{\epsilon\over 2} \cos^2\theta
\label{fourteen}
\end{equation}
which is of the same form as (\ref{eight}) (after the reparametrization of the Bloch sphere which
exchanges $x$ and $z$), and yields the same results (\ref{nine}) and (\ref{ten}) for the probability
distribution and average value of the fidelity, only with $N$ replaced by $4N$ everywhere. 

\section{Discussion}

From the analysis in the previous sections one can conclude that, if each pulse used for ``bang-bang''
control has a r.m.s. error of $\Delta$ radians in either the phase or the (integrated) amplitude, in
order to avoid loss of fidelity one must (roughly speaking) restrict the total number of cycles to
\begin{equation}
N_{max} \le {1\over \Delta^2}
\label{fifteen}
\end{equation}

In practice, of course, ``bang-bang'' control would be applied to a system which has an internal
Hamiltonian and is, as well, interacting with other systems, so the overall picture will be rather more
complicated than that described by Eq.~(\ref{one}), but there are grounds to believe that the conclusion
above should still hold true, approximately, in spite of these complications.  For instance, recent
calculations of decoherence due to internal interactions in a quantum computer show that the net effect
does not depend much on whether the computer is actually carrying out a calculation or merely sitting
idle in a predetermined state \cite{myself}.  Similarly, Miquel et al. \cite{miquel} found, in their
simulation of a
quantum computation with phase drift errors, that the effect of said errors could be well approximated by
a formula which did not depend on the details of the computation, only on the number of pulses.

As a simple test, I show in Fig.~3 the results of a simulation for a system with the self-Hamiltonian
\begin{equation}
H_0 = \hbar\Omega \sigma_z
\label{sixteen}
\end{equation}
initially prepared in the state $\ket{\psi_0} = {1\over{\sqrt 2}}(\ket+ + i \ket-)$, which, without
``bang-bang'' control, would evolve in time as $\ket{\psi_t} =  {1\over{\sqrt 2}}(e^{-i\Omega t}\ket+ + i
e^{i\Omega t} \ket-)$, and hence exhibit a fidelity ${\cal F} = \left|\bra{\psi_0}
U_{total}\ket{\psi_0}\right|^2 = \cos^2\Omega t$ (solid curve).  Low-noise ``bang-bang'' control (dashed
curve, amplitude noise, $N_{max}\Delta^2 = 0.1$) cancels out this free evolution quite effectively, so
the state fidelity remains close to 1 for the times shown, but if the noise is high instead (dotted
curve, $N_{max}\Delta^2 = 10$), the fidelity is completely lost.  This simulation assumes that the
control pulses are essentially instantaneous and very strong, so that the self-Hamiltonian
(\ref{sixteen}) can be ignored during a control pulse.  Also, in order to keep the fidelity close to 1,
the interval between pulses $\Delta t$ needs to be short enough; specifically, in this case, shorter than
$2\pi/\Omega$.  For Figure 3, $\Delta t = 0.005 \pi/\Omega$.

It should be pointed out that the initial state
chosen for the simulation in Fig.~3 (an eigenstate of $\sigma_y$) is just about the ``worst'' possible
choice, that is, it is one of the states that minimize the fidelity as given by Eq.~(\ref{eight}) (the
average fidelity for such states is given by Eq.~(\ref{tenandahalf})).  This explains why the simulations
in Fig.~3 look somewhat worse than those in Fig.~2. 
Nonetheless, this initial state was chosen because it gives maximum modulation of the fidelity under the
self-Hamiltonian
(\ref{sixteen}).

In Ref. \cite{viola1} it was shown that, in order to combat decoherence by the ``bang-bang''
method, one needs to have at least one complete cycle every $\tau_c$, where $\tau_c$ is the
characteristic ``coherence time'' of the environment.  Putting this condition together with
(\ref{fifteen}) yields the result that the ``bang-bang'' method with imperfect pulses can protect a
quantum information storing (or processing) device for a time no longer than
\begin{equation}
T_{max} < {\tau_c\over \Delta^2}
\label{seventeen}
\end{equation}
where $\Delta$ is the r.m.s. error per pulse.  For instance, if the $\pi$ pulses are known to be accurate
to, say, one part in $10^5$, one may expect $\Delta \sim \pi\times 10^{-5}$, and $T_{max} \sim 10^9
\tau_c$  

It is important to keep in mind that $\tau_c$ is not, in general, equal to the decoherence time of the
system itself under its interaction with the environment.  As an extreme example, the spontaneous
emission of an atom has a decoherence time of the order of $1/\gamma$, where $\gamma$ may be mega- or
gigahertz for an optical transition, but the environment responsible for the decoherence---namely, the
quantized electromagnetic field---has a ``coherence time'' of the order of an optical period or shorter,
that is, about $10^{-15}$ s.  On the other hand, for other kinds of decoherence---such as, for instance,
the kinds resulting from internal interactions between the qubits---one may expect the relevant $\tau_c$
to be of the order of a characteristic interaction frequency, which would also determine the basic
decoherence timescale.  In that case, ``bang-bang'' control with accurate pulses could substantially
lengthen the decoherence time, as Eq.~(\ref{seventeen}) indicates.

Discussions with C.-P. Yang are gratefully acknowledged.  This work has been supported in part by the
National Security Agency (NSA)
and Advanced Research and Development Activity (ARDA) under Army Research
Office (ARO) contract number DAAD19-99-1-0118; and by the National Science
Foundation under grant PHY-9802413.

\begin{figure}
\caption{Probability distribution of the fidelity for a totally random (uniformly distributed on the
Bloch sphere) initial state, and Gaussian noise of standard deviation $\Delta$ per pulse, for different
values of $ N \Delta^2$.  Solid line:  $N \Delta^2=0.1$.  Dashed line: $N \Delta^2=1$.  Dotted line: $N
\Delta^2=10$}
\end{figure}

\begin{figure}
\caption{Fidelity as a function of number of pulses for different values of $\Delta$.  Solid line: 
$N_{max} \Delta^2=0.1$.  Dashed line: $N_{max} \Delta^2=1$.  Dotted line:  $N_{max} \Delta^2=10$.  In
all cases $N_{max} = 400$.}
\end{figure}

\begin{figure}
\caption{Fidelity as a function of number of pulses for a system with the self-Hamiltonian
(\ref{sixteen}) for different values of $\Delta$.  Solid line: free evolution, no pulses. Dashed line:
$N_{max} \Delta^2=0.1$.  Dotted line:  $N_{max} \Delta^2=10$.  In
all cases $N_{max} = 400$.}
\end{figure}


\begin{references} 
\bibitem{viola1} Viola, L., and Lloyd, S. , 1998, {\it Phys.\ Rev.\ A} {\bf 58}, 2733.
\bibitem{ban} Ban, M., 1998 {\it J.\ Mod.\ Opt.} {\bf 45}, 2315.
\bibitem{viola2} Viola, L., Knill, E., and Lloyd, S., 1999, {\it Phys.\ Rev.\ Lett.} {\bf 82}, 2417; 
Viola, L., Lloyd, S., and Knill, E., 1999, {\it Phys.\ Rev.\ Lett.} {\bf 83},Ê4888, Viola, L., Knill, E.,
and Lloyd, S., 2000, LANL e-print quant-ph/0002072.
\bibitem{vitali} Vitali, D., and Tombesi, P., 1999 {\it Phys.\ Rev.\ A} {\bf 59}, 4178.
\bibitem{agarwal} Agarwal, G. S., 2000, {\it Phys.\ Rev.\ A} {\bf 61}, 013809.
\bibitem{duan} Duan, L.-M., and Guo, G.-C., 1998, {\it Phys.\ Rev.\ A} {\bf 57}, 2399.
\bibitem{kwiat} Berglund, A. J., Kwiat, P. G., and White, A. G., in preparation.
\bibitem{myself} Gea-Banacloche, J., 1999, {\it Phys.\ Rev.\ A} {\bf 60}, 185.
\bibitem{miquel}  Miquel, C., Paz, J. P., and Zurek, W. H., 1997, {\it Phys.\ Rev.\ Lett.} {\bf 78},
3971.
\end{references}
\end{document}